\documentclass[aps,prb,showpacs,reprint]{revtex4-1}

\usepackage{graphicx}

\begin{document}
\title
{Intrinsic oscillations of spin current polarization in a paramagnetic resonant tunneling diode}
\author{P. W\'ojcik}
\author{J. Adamowski}
\email[Electronic address: ]{adamowski@fis.agh.edu.pl}
\author{M. Wo{\l}oszyn}
\author{B.J. Spisak}
\affiliation{AGH University of Science and Technology, Faculty of
Physics and Applied Computer Science, al.~A.~Mickiewicza 30,
30-059~Krak\'ow, Poland}

\begin {abstract}
A spin- and time-dependent electron transport has been studied in
a paramagnetic resonant tunneling diode using the self-consistent
Wigner-Poisson method.  Based on the calculated current-voltage
characteristics in an external magnetic field
we have demonstrated that under a constant bias both the spin-up
and spin-down current components exhibit the THz
oscillations in two different bias voltage regimes. We have shown
that the oscillations of the spin-up (down) polarized current
result from the coupling between the two resonance states: one
localized in the triangular quantum well created in the emitter
region and the second localized in the main quantum well. We have
also elaborated the one-electron model of the current
oscillations, which confirms the results obtained with the
Wigner-Poisson method. The spin current oscillations can lower the
effectiveness of spin filters based on the paramagnetic resonant
tunneling structures and can be used to design the generators of
the spin polarized current THz oscillations that can operate under
the steady bias and constant magnetic field.%
\footnote{To be published in Phys.Rev.B.\\
Copyright (2012) by the American Physical Society.}
\end{abstract}


\maketitle

\section{Introduction}\label{sec:intro}

Dilute magnetic semiconductors (DMSs) such as ZnMnSe and GaMnAs
are promising materials due to their potential applications
in spintronics.\cite{Furdyna1988,Dietl2001,Sandratskii2003,Ando2003,Liu2005,Jungwirth2006,Maksimov2010}
The recent homo- and heteroepitaxy
methods\cite{Koo2010,Agarwal2006,Edmonds2002,Yu2002,Chiba2003,Ku2003,Eid2005,Jungwirth2005}
allow to deposit the DMS layers with thicknesses of
few nanometers, which enables to fabricate spintronic
nanodevices with the spin polarization of the current being
controlled by the external magnetic and electric fields.

The spin filter based on the DMS was proposed by Euges.\cite{Egues1998}
The nanostructure studied in Ref.~\onlinecite{Egues1998} consisted of a paramagnetic
semiconductor layer made from Zn$_{1-x}$Mn$_x$Se sandwiched between two non-magnetic
ZnSe layers.  The external magnetic field leads to the giant Zeeman splitting
of the conduction band minima in the paramagnetic layer, which causes that this layer acts as
a potential well for spin-down electrons and a potential barrier for spin-up electrons.
As a result, the total current flowing through the nanostructure is
dominated by the spin-down electrons.
In such a spin filter,\cite{Egues1998} the change of the spin polarization of the current
requires the change of the external magnetic field.
In the recently fabricated paramagnetic resonant tunneling diodes (RTDs)\cite{Slobodskyy2003,Ruth2011,Ohya2005},
the electrical control of the spin polarization of the current has been achieved.
The spin-polarized current is controlled by the bias voltage in
the presence of the external magnetic field
in paramagnetic RTDs\cite{Slobodskyy2003,Beletskii2005,Saffarzadeh2009,Ruth2011}
or even without the external magnetic field in ferromagnetic RTDs.\cite{Li2009,Wang2009,Qui2008,Li2007,Li2006, Ertler2006}
If the quantum well in the RTD is made from the DMS, the spin splitting
of the quasi-bound state energy level gives rise to the resonant tunneling conditions for
the spin-up and spin-down electrons satisfied for different bias voltages.
This leads to the separation of both the spin current components
and consequently to the spin polarization of the net current.
The operation of the paramagnetic RTD based on ZnSe/ZnBeSe/ZnMnSe heterostructure
as a spin filter has been experimentally demonstrated by
Slobodskyy et al.\cite{Slobodskyy2003} and theoretically described by Havu et al.\cite{Havu2005}
Recently, the spin polarization of the current in the paramagnetic RTD has been reported
at zero magnetic field.\cite{Ruth2011}
All these studies of the spin-polarized currents in the magnetic RTDs were devoted to
the stationary (steady) currents.
The oscillations of the spin-polarized currents in magnetic RTDs
have not been studied until now, although the intrinsic oscillations of the current have been detected
in the non-magnetic resonant tunneling structures.\cite{Orihashi2005,Kishimoto2007,Suzuki2009}

The intrinsic current oscillations occurring in the non-magnetic RTDs are intensively studied
due to their potential application as THz generators.
Recently, the THz oscillations have been experimentally observed in the GaInAs/AlAs
RTD integrated with a slot antenna.\cite{Orihashi2005,Kishimoto2007,Suzuki2009}
The attempts undertaken in order to explain the origin of these oscillations
led to contradictory conclusions.
Ricco and Azbel\cite{Ricco1984} argued that the oscillations of the current
in the non-magnetic RTD result from the fact that the system enters and leaves the resonant
current conditions.
According to this model,\cite{Ricco1984} the intrinsic oscillations of the current should occur at the resonance bias
that is in contradiction with the experimental\cite{Orihashi2005,Kishimoto2007,Suzuki2009}
and numerical\cite{Jensen1991} studies, which show that the current oscillations occur only
in the negative differential resistance (NDR) regime of the current-voltage characteristics, i.e., above the resonance bias.
Woolard~et~al.\cite{Woolard1996} suggested that the current oscillations in the non-magnetic RTD result from
the charge fluctuations in the potential well created between the emitter and the nearest barrier.
This proposition was extended by Zhao~et~al.,\cite{Zhao2000,Zhao2001,Zhao2003} who
showed that the intrinsic oscillations are due to the coupling between
the quasi-bound state localized in the emitter-related quantum well
and the quasi-bound state in the main quantum well.
However, this model\cite{Zhao2000,Zhao2001,Zhao2003}
is based on the adiabatic approximation, according to which the electron states
are slowly varying in time. This assumption is not valid in the THz oscillation regime.
Moreover, the explanation given by Zhao~et~al.\cite{Zhao2000,Zhao2001,Zhao2003}
does not answer the question, why the current oscillations do not decay in time as a result
of the dissipative factor corresponding to the imaginary part of the resonance state energy.
In our recent paper,\cite{Wojcik2010} we have shown that the intrinsic oscillations
of the spin-unpolarized current result from the coupling between the two arbitrary quasi-bound states in the
non-magnetic triple-barrier resonant tunneling structure.  We have found the intrinsic oscillations in two bias ranges.
We have interpreted the oscillations that occur at the bias values below the resonance bias as resulting
from the coupling between the quasi-bound states in both the quantum wells, while
the oscillations that occur in the NDR regime as resulting from the coupling
between the emitter-related quasi-bound state and the quasi-bound state localized in the nearby quantum well.

In the present paper, we study the spin- and time-dependent electronic
transport in the paramagnetic double-barrier RTD based on ZnSe/ZnBeSe/ZnMnSe.
We show that -- under certain well-defined constant bias and external magnetic field
-- the intrinsic oscillations of both the spin current components
appear, which leads to the oscillations of the spin polarization of the total current.
Based on the analysis of the spin- and time-dependent potential energy profiles and electron density distributions,
we demonstrate that the oscillations of the spin polarized current result from the coupling between the
quasi-bound state localized in the triangular quantum well created in the emitter (E) region
with the spin-dependent quasi-bound state localized in the main quantum well (QW).
The present results show that at certain bias voltages the spin polarization of the current
flowing through the paramagnetic RTD is not constant but oscillates with the THz frequency.
This effect can be of crucial importance for the possible applications of this
nanodevice as a spin filter and THz generator.

The paper is organized as follows: in Sec. II, we describe the theoretical model of the paramagnetic RTD
and the time-dependent Wigner-Poisson method.  Section III contains
the results, Section IV -- the discussion, and Section V -- conclusions and summary.

\section{Theoretical model}\label{sec:model}

\begin{figure}[htbp]
\includegraphics[width=0.9\columnwidth]{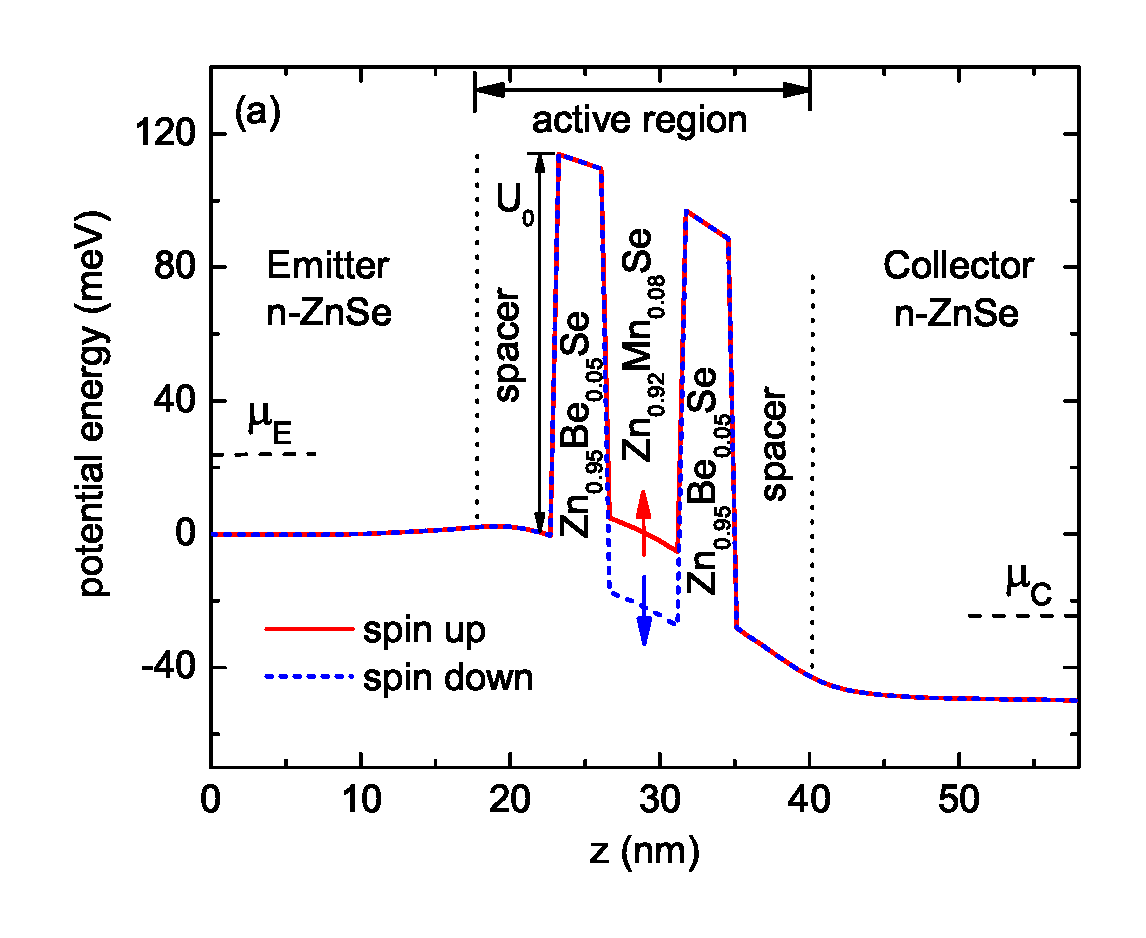}
\caption{\label{fig:1}(Color online) Potential energy in the
paramagnetic RTD for the electrons with spin-up (solid, red) and spin-down (dotted, blue).
Coordinate $z$ is measured along the growth direction,
$\mu_{E(C)}$ is the electrochemical potential of the emitter (collector).
The active (undoped)  region consists of the paramagnetic quantum well
made from Zn$_{0.92}$Mn$_{0.08}$Se sandwiched between the two
Zn$_{0.95}$Be$_{0.05}$Se potential barriers
and is separated from the $n$-doped ZnSe ohmic contacts by the two ZnSe spacer layers.}
\end{figure}

We consider the spin-dependent electronic transport through the paramagnetic RTD that consists of the paramagnetic
quantum-well layer made from Zn$_{1-x}$Mn$_{x}$Se embedded between the two barrier layers made
from Zn$_{0.95}$Be$_{0.05}$Se (Fig.~\ref{fig:1}).
The active (undoped) region of the nanodevice is separated from the $n$-doped ZnSe
ohmic contacts by the two spacer layers made from ZnSe.
In the presence of the external magnetic field $\mathbf{B} = (0,0,B)$ applied in the growth ($z$)
direction, the exchange interaction between the conduction band
electrons and the Mn$^{2+}$ ions leads to the giant Zeeman splitting  of the conduction band
minimum in the paramagnetic quantum-well layer.\cite{Furdyna1988}
In the external magnetic field, the conduction-band electrons form the Landau states
with the wave functions spread over the $x-y$ plane.
Therefore, the one-electron problem can be separated into $(x, y)$ and $z$ coordinates and
the electronic transport between the emitter and collector can be
described as the one-dimensional motion of the electron in the $z$ direction.

In order to simulate the electronic transport through the nanostructure,
we apply the time-dependent Wigner-Poisson approach, according to which
the conduction band electrons are described by the spin-dependent Wigner distribution
function.\cite{Spisak2009,Weng2003,Spisak2005}
The  quantum kinetic equation takes on the following form\cite{Spisak2009}
\begin{eqnarray}
\label{eq:Wequation}
\frac{\partial \rho^\mathcal{W}_{\sigma}(z,k,t)}{\partial t} &+&
\frac{\hbar k}{m}\frac{\partial \rho^\mathcal{W}_{\sigma}(z,k,t)}{\partial z} \\ \nonumber
 &=& \frac{i}{2\pi \hbar}\int\limits_{-\infty}^{+\infty}\!{\mathrm{d}k^{\prime}}\:
 \mathcal{U}_{\sigma}(z,k-k^{\prime},t) \rho^\mathcal{W}_{\sigma}(z, k^{\prime},t) \;,
\end{eqnarray}
where $\rho^\mathcal{W}_{\sigma}(z,k,t)$ is the spin-dependent Wigner distribution function, $k$ is the
$z$-component of the wave vector, $m$ is the electron conduction-band effective
mass, and $\sigma=(\uparrow, \downarrow)$ is the spin index.

The non-local potential $\mathcal{U}_{\sigma}(z,k-k^{\prime};t)$
for spin channel $\sigma$  is given by the formula
\begin{eqnarray}
 \mathcal{U}_{\sigma}(z,k-k^{\prime},t) & = &
\int\limits_{-\infty}^{+\infty}\!{\mathrm{d}z^{\prime}}~\big[U_{\sigma}(z+z^{\prime}/2,t)
\label{eq:npot}\\*
 & - & U_{\sigma}(z-z^{\prime}/2,t)\big]
 \exp{\big[-i(k-k^{\prime})z^{\prime}\big]} \;, \nonumber
\end{eqnarray}
where $U_{\sigma}(z,t)$ is the spin-dependent potential energy,
which can be expressed as the sum of the two terms
\begin {equation}
U_{\sigma}(z,t)=U_{\sigma}^{cb}(z,B)+U_{el}(z,t). \label{Us}
\end {equation}
In Eq.~(\ref{Us}), the first term denotes the spin-dependent
potential energy of the conduction-band bottom, while the second
term is the potential energy of an electron in the electric field
acting in the nanostructure and has the form
\begin{equation}
U_{el}(z,t) = U_{V_b}(z) + U_H(z,t) \;,
\label{Uel}
\end{equation}
where $U_{V_b}(z)$ is the potential energy of an electron in the electric field
generated by voltage $V_b$ applied between the emitter and collector
and $U_H(z;t)$ is the Hartree energy that takes into account the electron-electron interactions.
In Eq.~(\ref{Us}), we have neglected the exchange energy (see Subsection IV.C).

The conduction-band potential energy profile has the form
\begin{equation}
U_{\sigma}^{cb}(z,B)=
\left \{
\begin{array}{cl}
U_0 \;, & \mbox{if $z_1 \leq z \leq z_2$ and $z_3 \leq z \leq z_4$},\\
\pm E_Z(B) \;, & \mbox{if $z_2 \leq z \leq z_3$},\\
0 \;, & \mbox{otherwise},
\end{array}
\right.
\end{equation}
where $z_1$ and $z_2$ ($z_3$ and $z_4$) are the positions of the left (right) potential barrier interfaces,
$U_0$ is the height of the potential barrier,
and $E_Z(B)$ is the Zeeman energy of an electron in the
paramagnetic quantum-well layer, which is positive (negative) for the spin-up (spin-down) electrons.
For a small concentration of Mn$^{2+}$ ions the Zeeman energy can be expressed by the
formula\cite{Furdyna1988}
\begin{equation}
E_Z (B)= \frac{1}{2} \, N_0\alpha \; x \, S_0 \,
\mathcal{B}_S \! \left( \frac{g \mu_B SB}{k_B T_{ef\!f}} \right ) \; ,
\label{eq:GZS}
\end{equation}
where $N_0\alpha=0.26$ eV is the {\it sp-d} exchange constant, $x$ is the concentration
of Mn$^{2+}$ ions,  $\mathcal{B}_S$ is the Brillouin function for spin
$S=5/2$ that corresponds to the spin of Mn$^{2+}$ ion, $g$ is the
effective Land{\'e} factor, $\mu _B$ is the Bohr magneton,
$S_0$ and $T_{eff}$ are the phenomenological parameters corresponding to the antiferromagnetic interaction
between the Mn$^{2+}$ ions and have been taken on as $S_0=1.18$ and $T_{\mathrm{eff}}=2.55$~K for $x=0.083$.\cite{Twardowski2005}

Potential energy $U_{el}(z,t)$ satisfies the Poisson equation
\begin{equation}
\frac{\mathrm{d}^2 U_{el}(z,t)}{\mathrm{d}z^2}=
\frac{e^2}{\varepsilon_0\varepsilon}[N_D(z)-n(z,t)] \;,
\label{eq:PEq}
\end{equation}
where $e$ is the elementary charge, $\varepsilon_0$ is the vacuum electric permittivity, $\varepsilon$
is the relative static electric permittivity, $N_D(z)$ is the
concentration of the ionized donors, and $n(z)=\sum_{\sigma} n_{\sigma}(z)$ is
the total density of electrons with $n_{\sigma}(z)$ being the density of the
electrons with spin $\sigma$.  The energy of the emitter conduction-band bottom
is taken as the reference energy and set equal to $0$.

In order to solve the system of non-linear equations (\ref{eq:Wequation}) and (\ref{eq:PEq})
we apply the self-consistent time-dependent procedure\cite{Gummel1964} with the boundary conditions
$U_{el}(0,t)=0$ and $U_{el}(L,t)=-eV_b$ for the Poisson equation,
where $L$ is the length of the nanodevice.
For the quantum kinetic equation (\ref{eq:Wequation}) we apply
the generalized form of the boundary conditions proposed by Frensley\cite{Frensley1990}
\begin{eqnarray}
\rho_{\sigma}^\mathcal{W}(0,k,t)\bigg|_{k>0}&=&f^E_{\sigma}(k) \; , \\ \nonumber
\rho_{\sigma}^\mathcal{W}(L,k,t)\bigg|_{k<0}&=&f^C_{\sigma}(k) \; .
\label{eq:bc}
\end{eqnarray}
Distribution function $f^{\nu}_{\sigma}(k)\; (\nu=E,C)$ is derived from the Fermi-Dirac distribution
function by summing over the Landau energy levels, which leads to
\begin{equation}
f^{\nu}_{\sigma}(k) = \frac{eB}{h}  \sum _{n=0}^{N_{max}}
\frac{1}{\exp{\left[\frac{1}{k_BT} \left (\frac{\hbar ^2 k^2}{2m}-E^{\nu}_{n\sigma} \right) \right] +1}} \;,
\label{eq:fd}
\end{equation}
where $T$ is the temperature,
$E^{\nu}_{n \sigma}=\mu_{\nu}-\hbar\omega _c\left ( n+ 1/2 \right ) - \sigma \mu _B B$,
$\mu_{\nu} = F_{\nu} - e V_{\nu}$ is the electrochemical potential of reservoir $\nu$,
$F_{\nu}$ is the corresponding Fermi energy, $V_{\nu}$ is the voltage applied to contact $\nu$,
and $\omega _c=eB/m$ is the cyclotron frequency.
In Eq.~(\ref{eq:fd}), $N_{max}$ determines the highest occupied Landau state and is determined by the density of
electrons in the reservoirs.

The spin-dependent Wigner distribution function allows us to calculate the spin-dependent electron density
\begin{equation}
n_{\sigma}(z,t)=\frac{1}{2\pi}\int\limits_{-\infty}^{+\infty}\!{\mathrm{d}k} \rho_{\sigma}^\mathcal{W}(z,k,t)
\label{eq: ed}
\end{equation}
and the spin-dependent current density
\begin{equation}
j_{\sigma}(t)=\frac{e}{2\pi L}\int\limits_0^L dz\int\limits_{-\infty}^{+\infty}\!{\mathrm{d}k}
\frac{\hbar k}{m}\rho_{\sigma}^\mathcal{W}(z,k,t) \;.
\label{eq:fc}
\end{equation}
The spin polarization of the current is defined as follows:
\begin{equation}
P(t)=\frac{j_{\uparrow}(t)-j_{\downarrow}(t)}{j_{\uparrow}(t)+j_{\downarrow}(t)} \:,
\label{eq:P}
\end{equation}
where $j_{\uparrow}$ and $j_{\downarrow}$ are
the spin-up and spin-down current densities, respectively.

\begin{figure*}[!htbp]
\includegraphics[width=1.3\columnwidth]{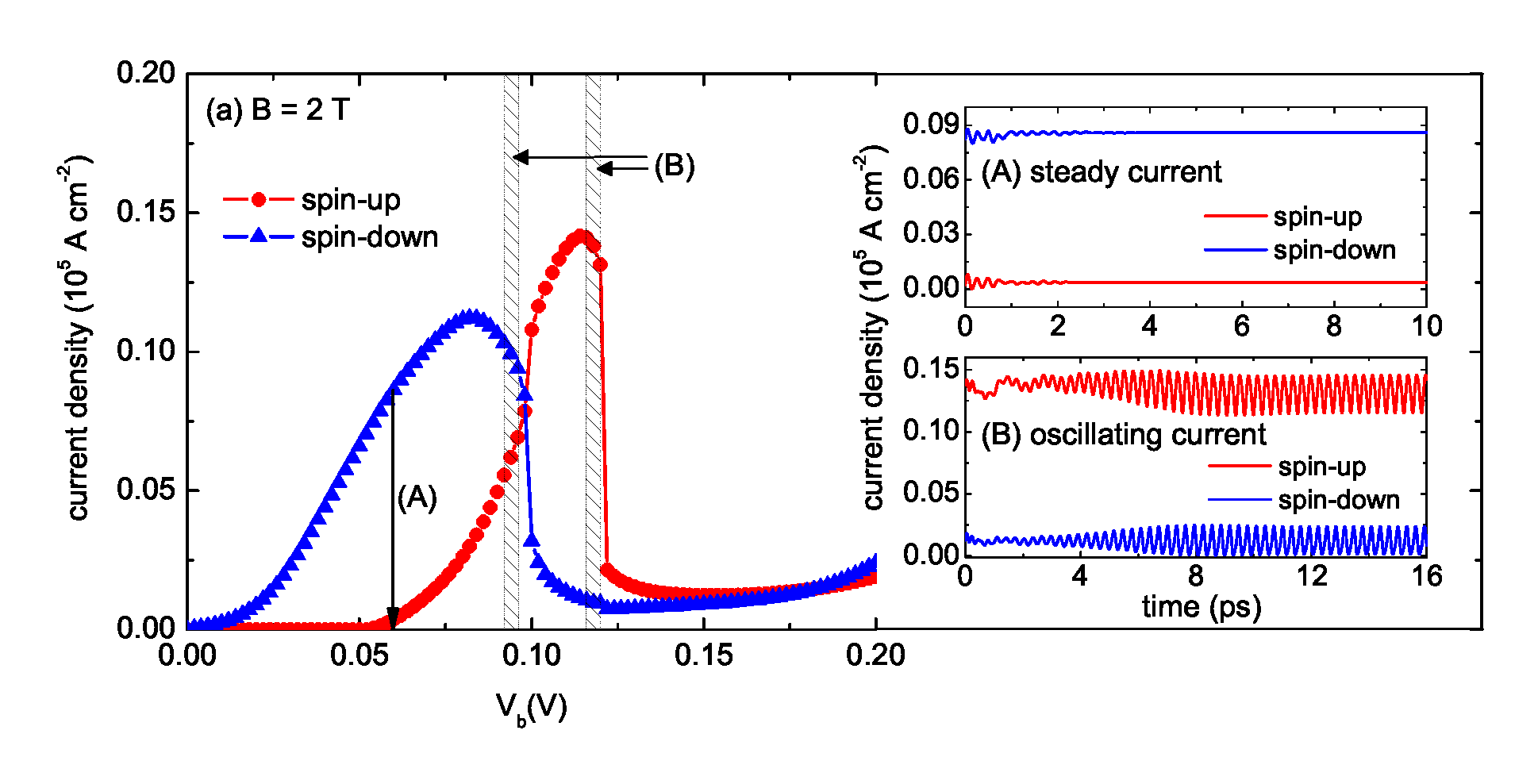}
\includegraphics[width=0.8\columnwidth]{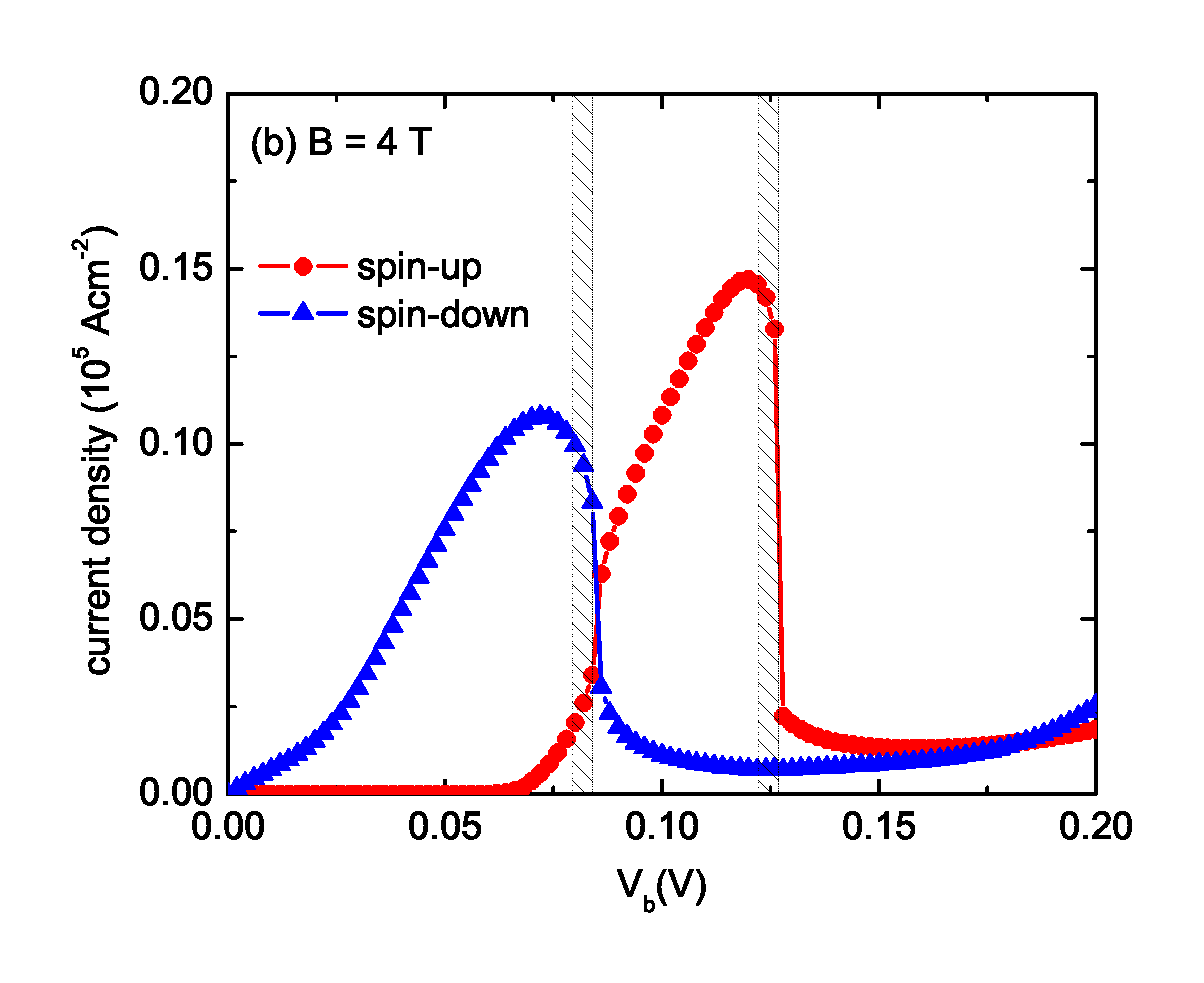}
\includegraphics[width=0.8\columnwidth]{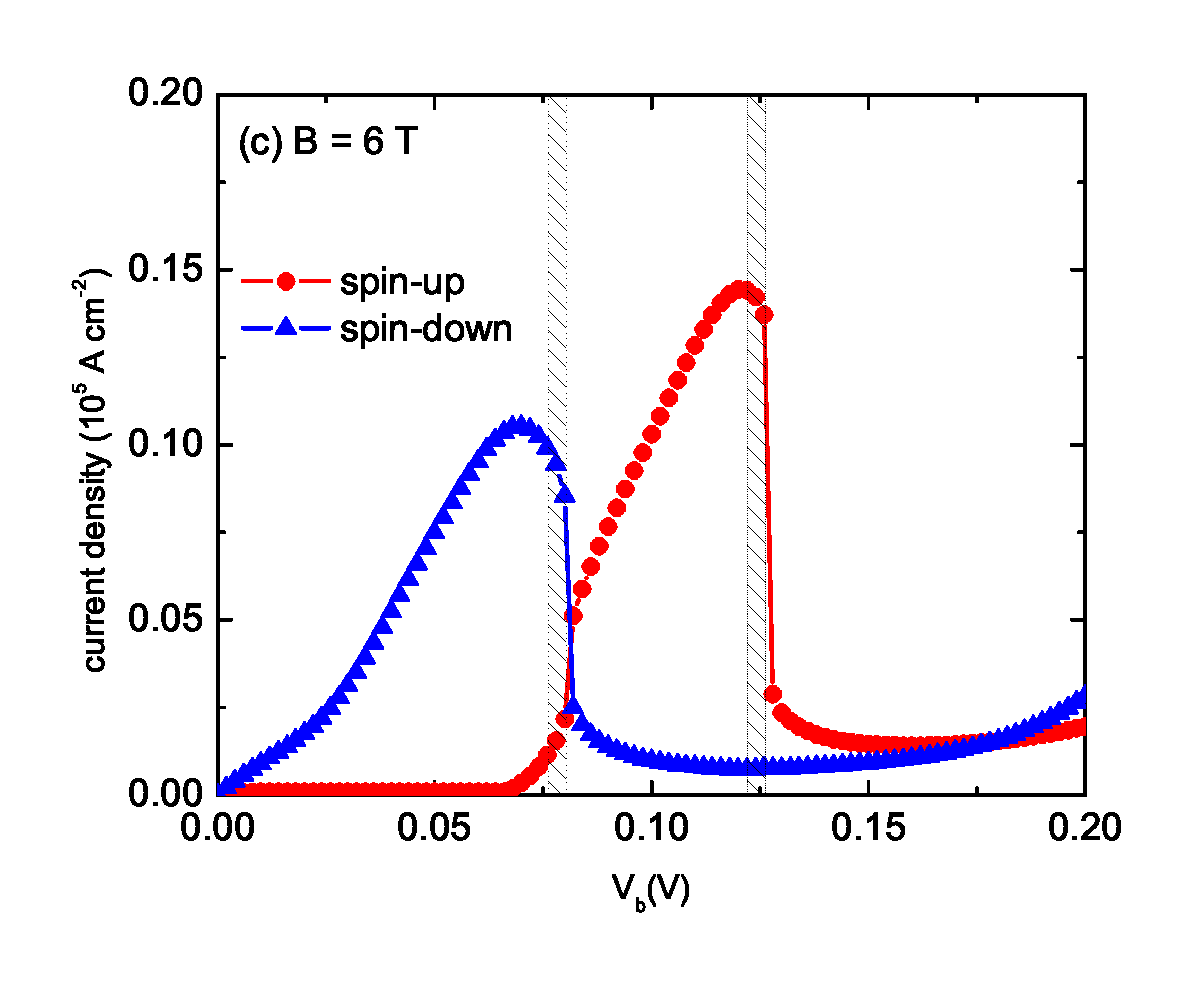}
\caption{\label{fig:2}(Color online) Current-voltage characteristics
for spin-up (curves with circles, red) and spin-down (curves with triangles, blue) electrons and for the magnetic field (a) $B=2$~T,
(b) $B=4$~T, and (c) $B=6$~T.
The hatched areas correspond to the bias regimes, in which the oscillations of the spin-polarized current occur.
The values of the current density in these regimes are calculated by averaging over one period of oscillations.
Inset (A) shows how the steady current is reached after several
initial fluctuations for the bias denoted by arrow (A) on main panel (a),
inset (B) shows the formation of the stable oscillations of the spin currents that occurs in the bias regimes denoted by
arrows (B) on main panel (a).}
\end{figure*}

In the present work, we focus on the oscillations of the spin-polarized currents.
In order to choose the nanostructure parameters, for which the current oscillations
are mostly pronounced, we have performed a number of computational runs that
allowed us to establish the optimal values of the parameters.
The present simulations have been performed for the following values of the nanostructure parameters:
the thickness of each contact is equal to 17~nm, the thickness of each spacer layer is 3~nm,
the thickness of the potential well layer is 5~nm, the total length of the nanodevice $L = 54$~nm,
and the Mn concentration $x=8.3$~\%.  The barriers are assumed to be symmetric with thickness 3 nm each.
We have taken on the height of both the potential barriers $U_0 = 0.115$~eV.\cite{Chauvet2000}
The contacts are made from the $n$-type ZnSe with the homogeneous  concentration of donors
$N_D=10^{18}$ cm$^{-3}$.
Due to the small thickness of the double-barrier region we assume
the effective electron mass of the ZnSe conduction band,
i.e., $m=0.16\,m_0$, where $m_0$ is the free electron mass.
We take on the relative electric permittivity of ZnSe $\varepsilon=8.6$ for the entire nanostructure.
The present simulations have been carried out for temperature $T=1.2$ K
on the computational grid with $N_z=95$ mesh points for coordinate $z$ and $N_k=72$ mesh points for $z$-component of the wave vector.
The $z$ coordinate increment was $\Delta z = a$, where $a = 0.5667$ nm is the lattice constant
of ZnSe.

\section{Results}\label{sec:results}
\begin{figure}[htbp]
\includegraphics[width=0.8\columnwidth]{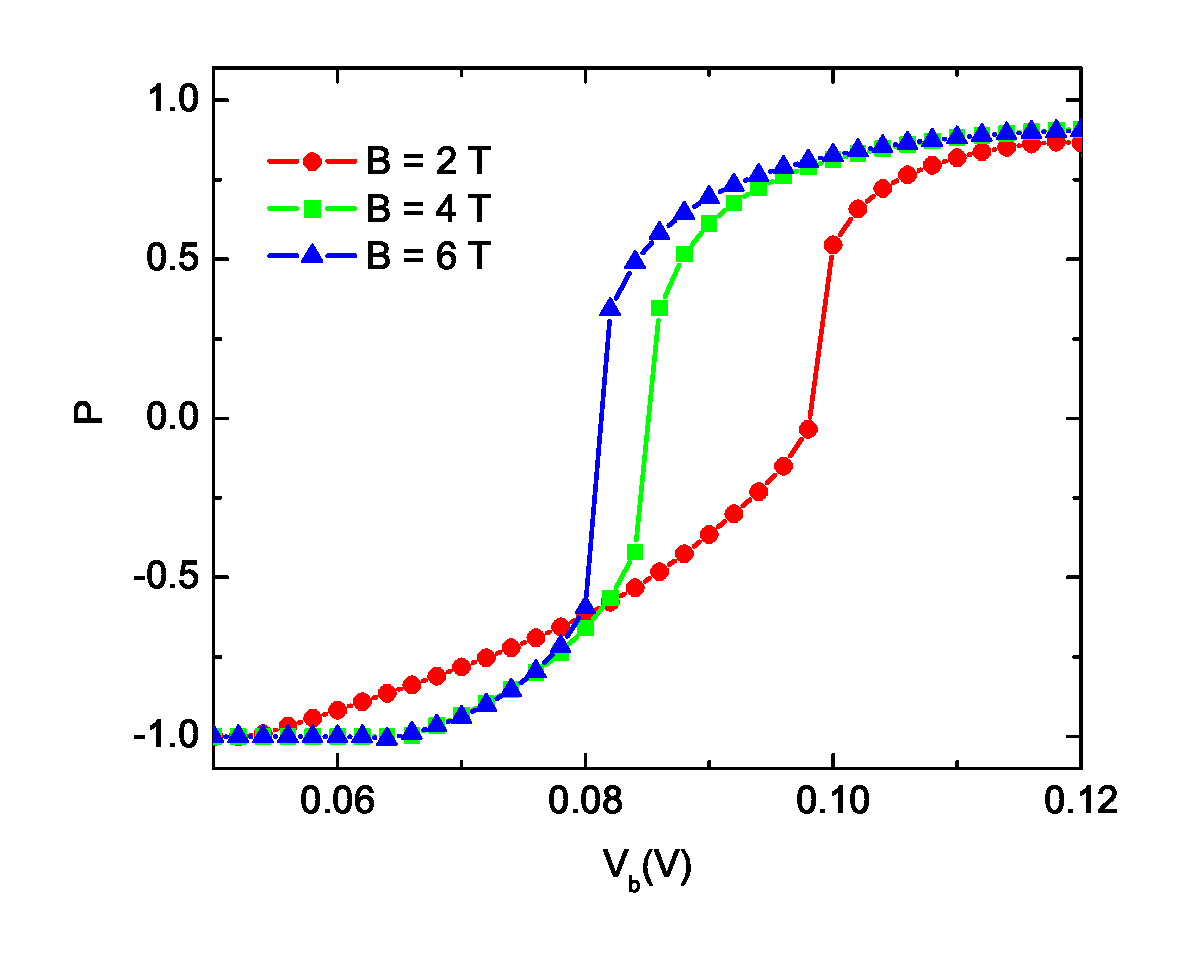}
\caption{\label{fig:3}(Color online)  Spin polarization $P$ of the current as a function
of bias voltage $V_b$ for the magnetic fields $B=2,4,6$~T.
For the oscillating current the spin polarization has been
calculated by averaging over one period of the oscillations.}
\end{figure}

The calculated current-voltage characteristics for spin-up and spin-down
components of the current in the presence of the external magnetic field are displayed on Fig.~\ref{fig:2}.
The resonant current peaks for both the spin current components are separated,
which results from the fact that the resonance conditions for the spin-up and spin-down electrons are
satisfied at different bias voltages due to the giant Zeeman splitting of the quasi-bound state energy levels
in the paramagnetic quantum well.
If the magnetic field increases the resonant current peak for the spin-down electrons shifts
towards the lower bias, while the resonant peak for the spin-up electrons
shifts towards the higher bias [Fig.~\ref{fig:2}(a,b,c)].
The separation of these peaks is the basic property that can be exploited
to fabricate an effective spin filter, in which
the spin polarization of the current is controlled by the bias voltage.
Figure~\ref{fig:3} shows the spin polarization of the current
as a function of the bias voltage for different magnetic fields.
For all values of the magnetic field the spin polarization of the current
varies from $P = -1$ for the low bias  to $P \simeq +1$ for the sufficiently high bias.
Moreover, for the higher magnetic field the transition between the almost fully spin
polarized currents occurs in a narrower bias interval.

The current-voltage characteristics presented in Fig.~\ref{fig:2} have been calculated
in the bias range from $V_b=0$ to $V_b=0.2$~V  with step $\delta V_b=0.005$~V.
For each step, the bias voltage has been changed to the next value by $\delta V_b$ if
both the spin current components reach the steady state as shown in
inset (A) of Fig.~\ref{fig:2}(a).
We have found that at certain bias voltages that belong to
the NDR regimes for the spin-up and spin-down electron currents both the spin currents
do not stabilize as the steady currents but oscillate with the THz frequency
[cf. inset (B) on Fig.~\ref{fig:2}(a)].

\begin{figure}[htbp]
\includegraphics[width=0.82\columnwidth]{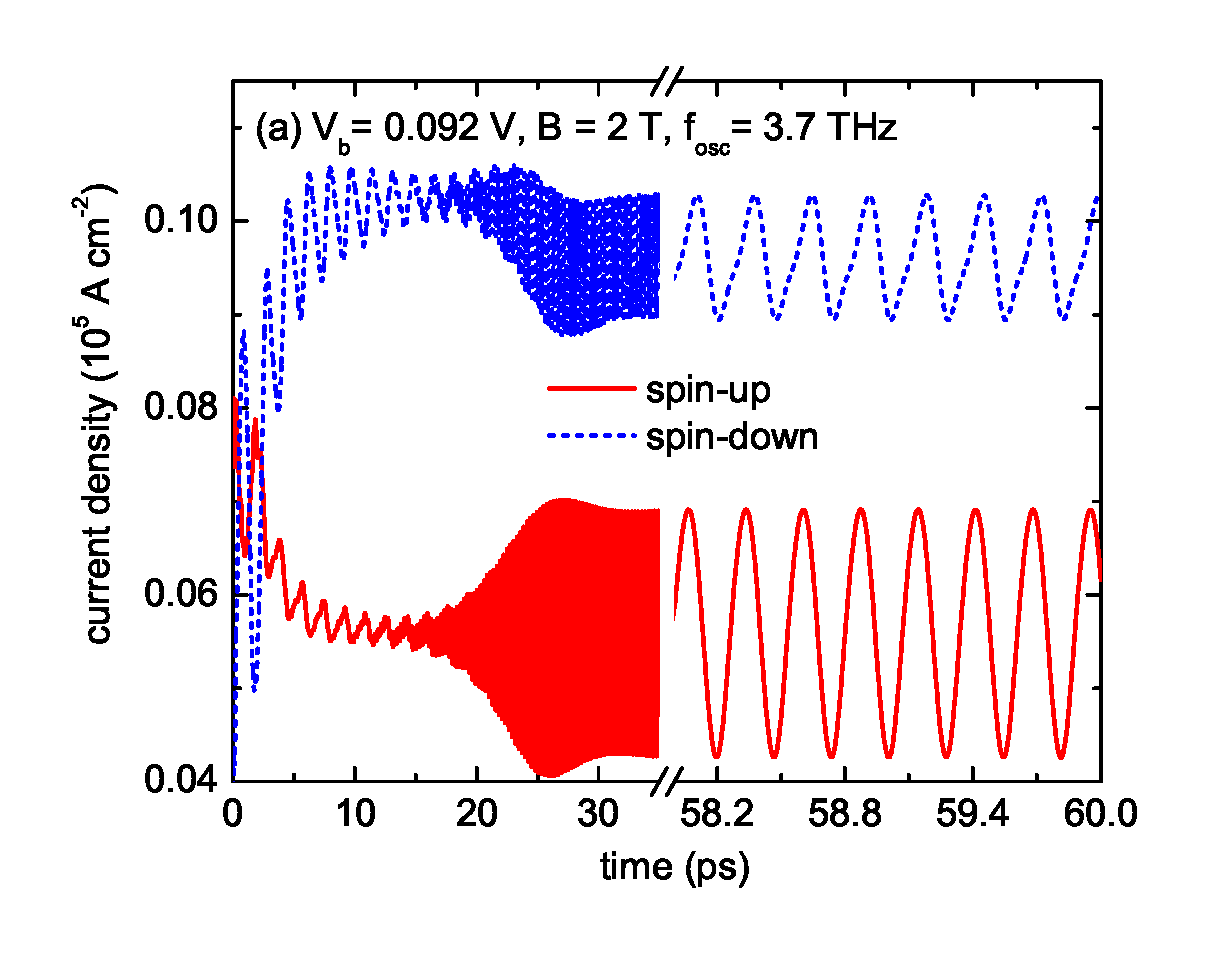}
\includegraphics[width=0.8\columnwidth]{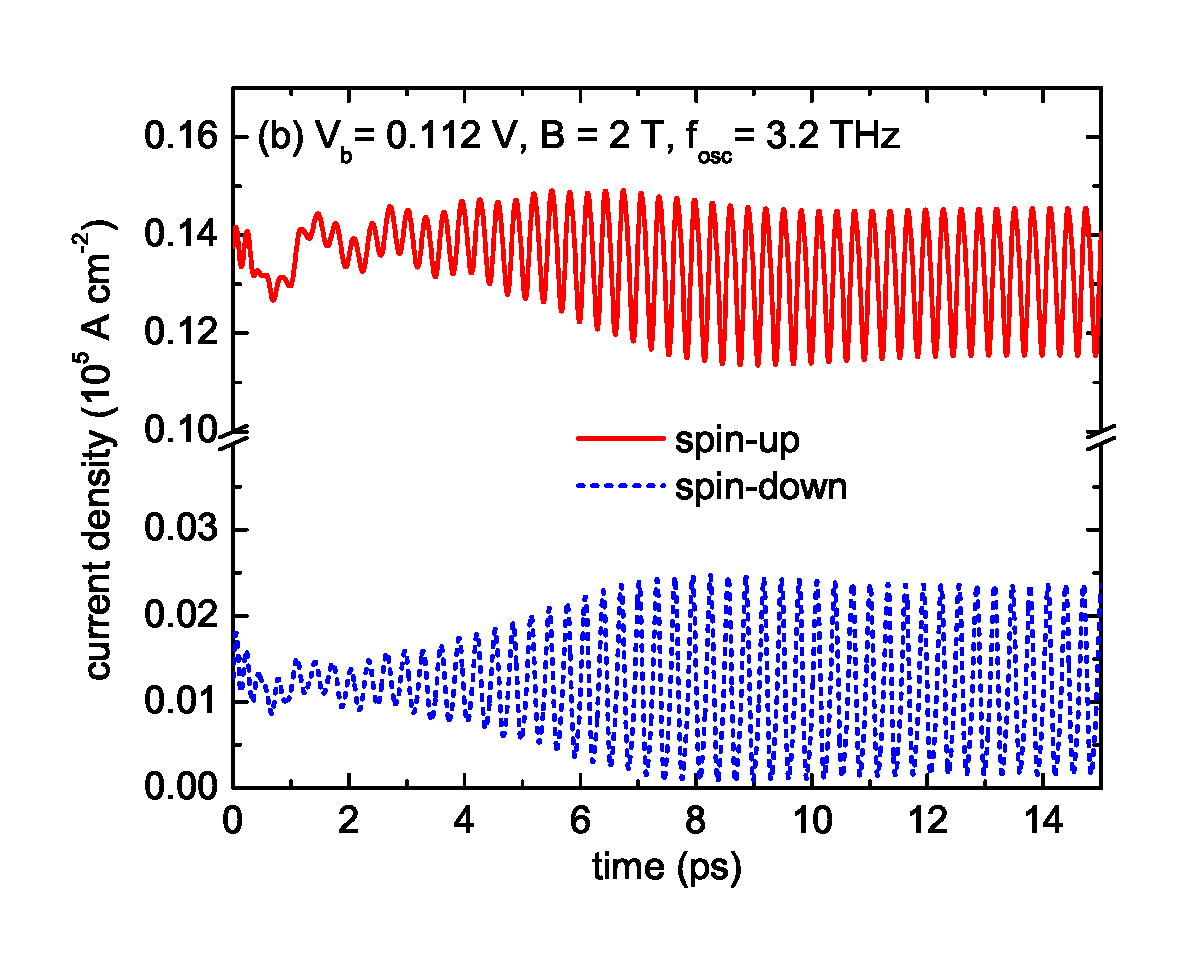}
\caption{\label{fig:4}(Color online)
Spin-up (solid, red) and spin-down (dashed, blue) current oscillations
in the NDR regimes for (a) the spin-down current component ($V_b=0.092$~V) and
(b) for the spin-up current component ($V_b=0.112$~V).
The calculations have been performed for $B=2$~T.}
\end{figure}

\begin{figure}[htbp]
\includegraphics[width=0.8\columnwidth]{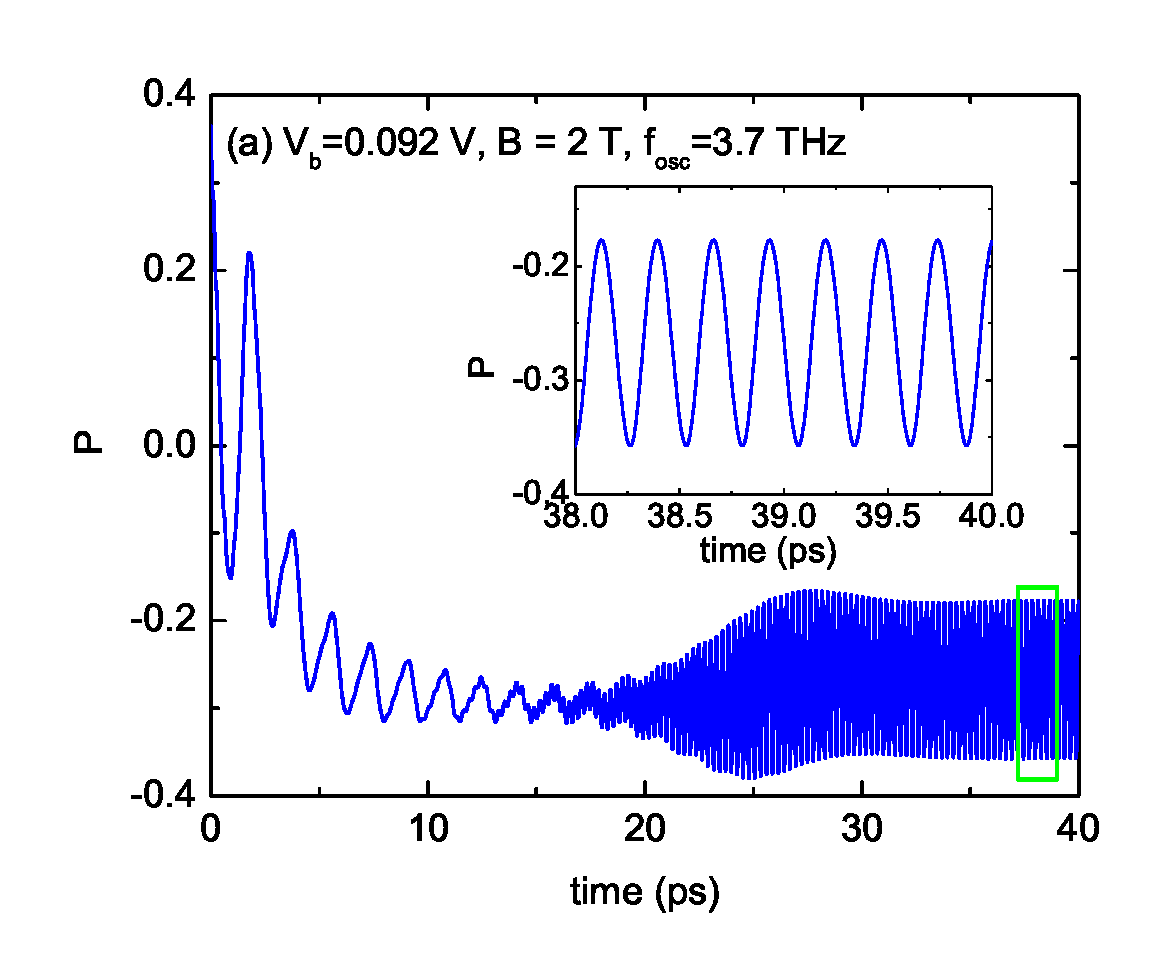}
\includegraphics[width=0.82\columnwidth]{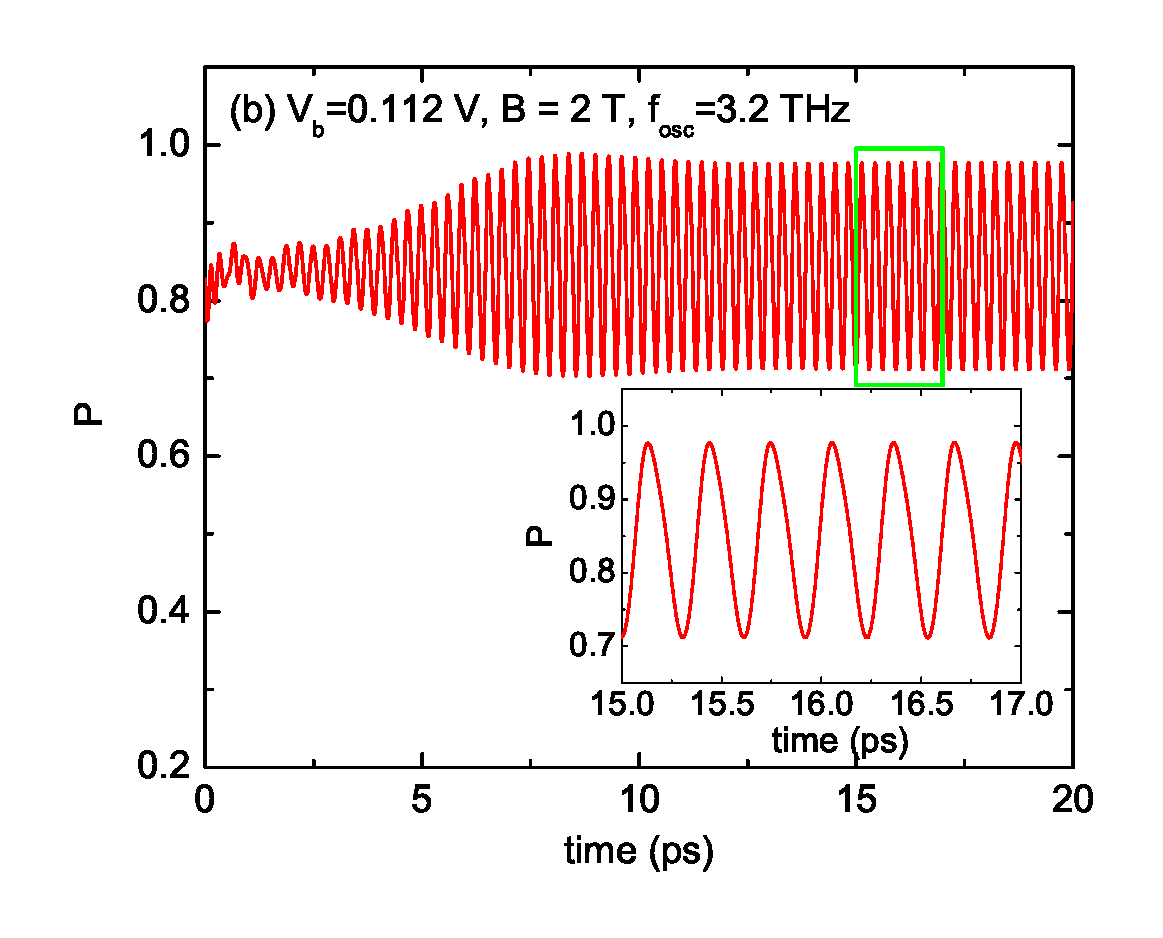}
\caption{\label{fig:5} (Color online) Oscillations of the spin current polarization $P$ in the NDR
regime for (a) the spin-down current component ($V_b=0.092$~V) and (b) spin-up current component ($V_b=0.112$~V).
Insets display the regions marked by the rectangles on the main panels.}
\end{figure}

In Fig.~\ref{fig:4}, we present the formation of the current oscillations
for the bias $V_b=0.092$~V that corresponds to the NDR regime for the spin-down current component [Fig.~\ref{fig:4}(a)]
and $V_b=0.112$~V that corresponds to the NDR regime for the spin-up current component [Fig.~\ref{fig:4}(b)].
In both the cases, the spin-polarized current oscillates with constant frequencies equal to
$f_{osc}=3.7$~THz for $V_b=0.092$~V and $f_{osc}=3.2$~THz for $V_b=0.112$~V.
For the  bias voltages, for which the current oscillations
have been found, the spin current polarization is not well defined but also
oscillates with the THz frequency.
Figure~\ref{fig:5} displays the oscillations of the spin current polarization
that occur in the NDR regime for the spin-down [Fig.~\ref{fig:5}(a)] and spin-up [Fig.~\ref{fig:5}(b)]
current component.
We note that the spin current polarization exhibits the oscillatory behavior in the bias regimes, in which
it reaches the maximal values.

\begin{figure}[htbp]
\includegraphics[width=0.8\columnwidth]{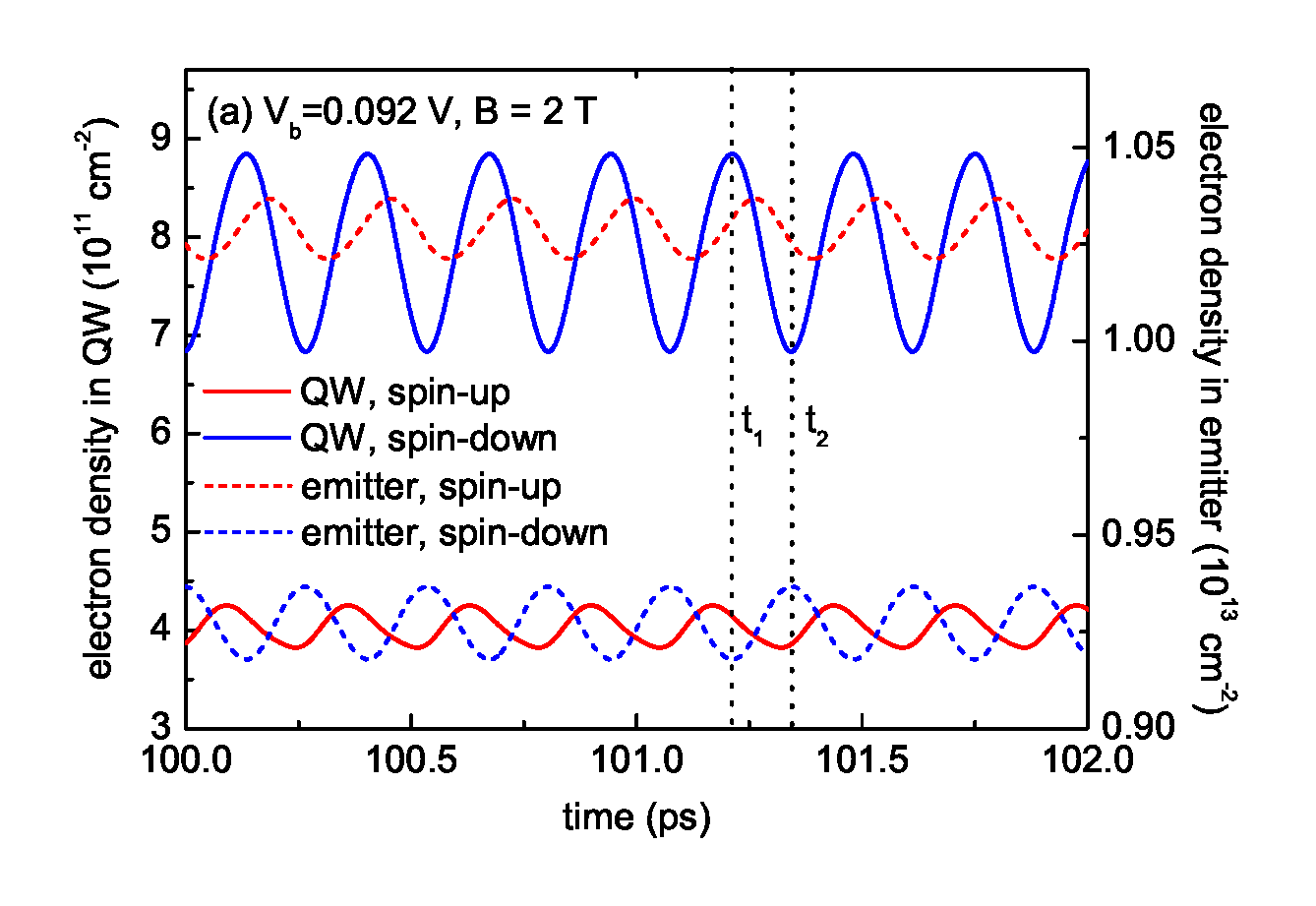}
\includegraphics[width=0.8\columnwidth]{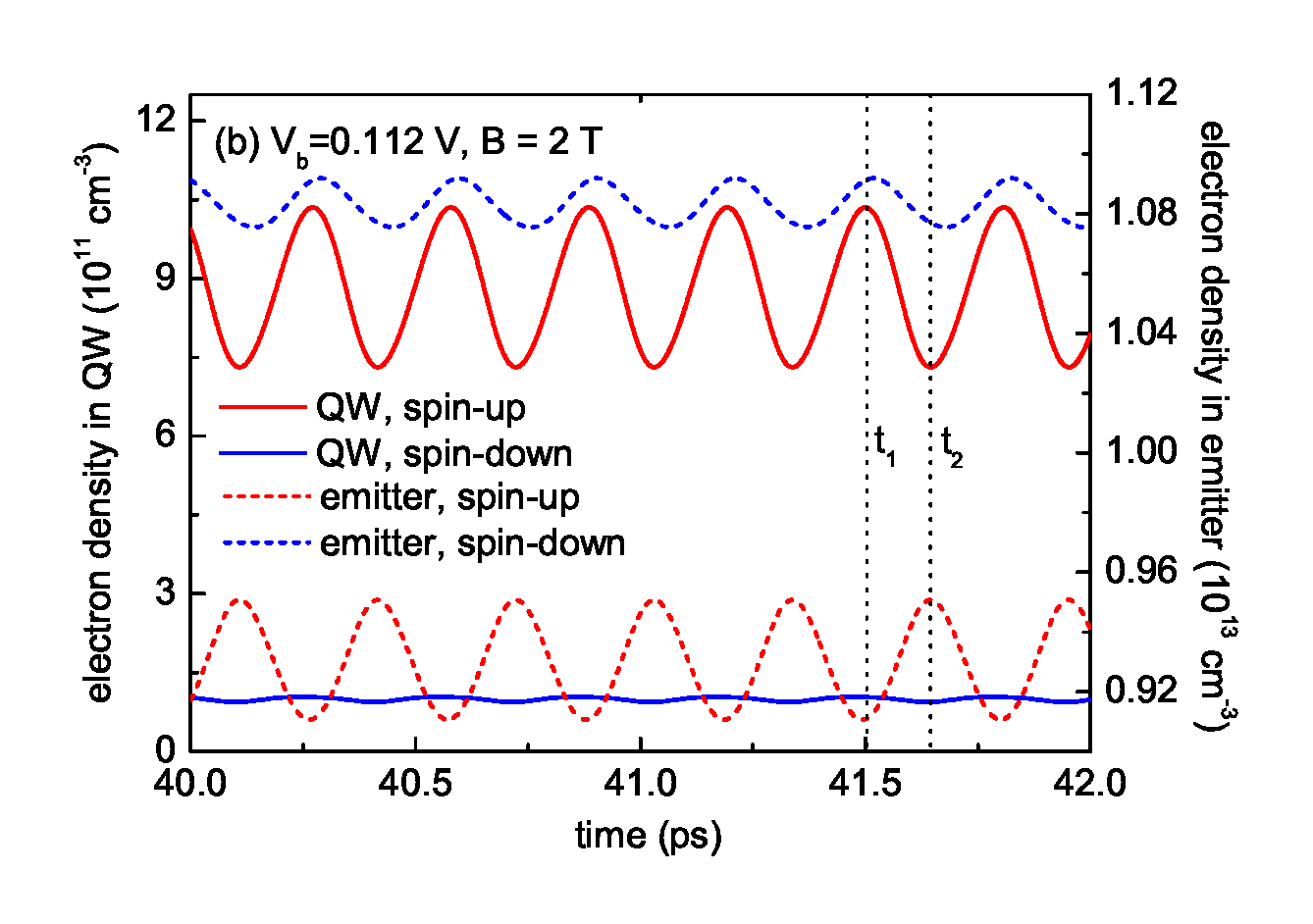}
\caption{\label{fig:6} (Color online) Oscillations of the spin-up (red) and spin-down (blue) density
of the electrons localized in the emitter region (dashed line, right scale) and the QW (solid line, left scale)
calculated for magnetic field $B=2$~T and the bias voltage from the NDR regimes for
(a) spin-down current component ($V_b=0.092$~V) and (b) spin-up current component ($V_b=0.112$~V).
The density of electrons localized in the QW reaches the maximum (minimum) at time instant $t_1$ ($t_2$).}
\end{figure}

\begin{figure}[htbp]
\includegraphics[width=0.8\columnwidth]{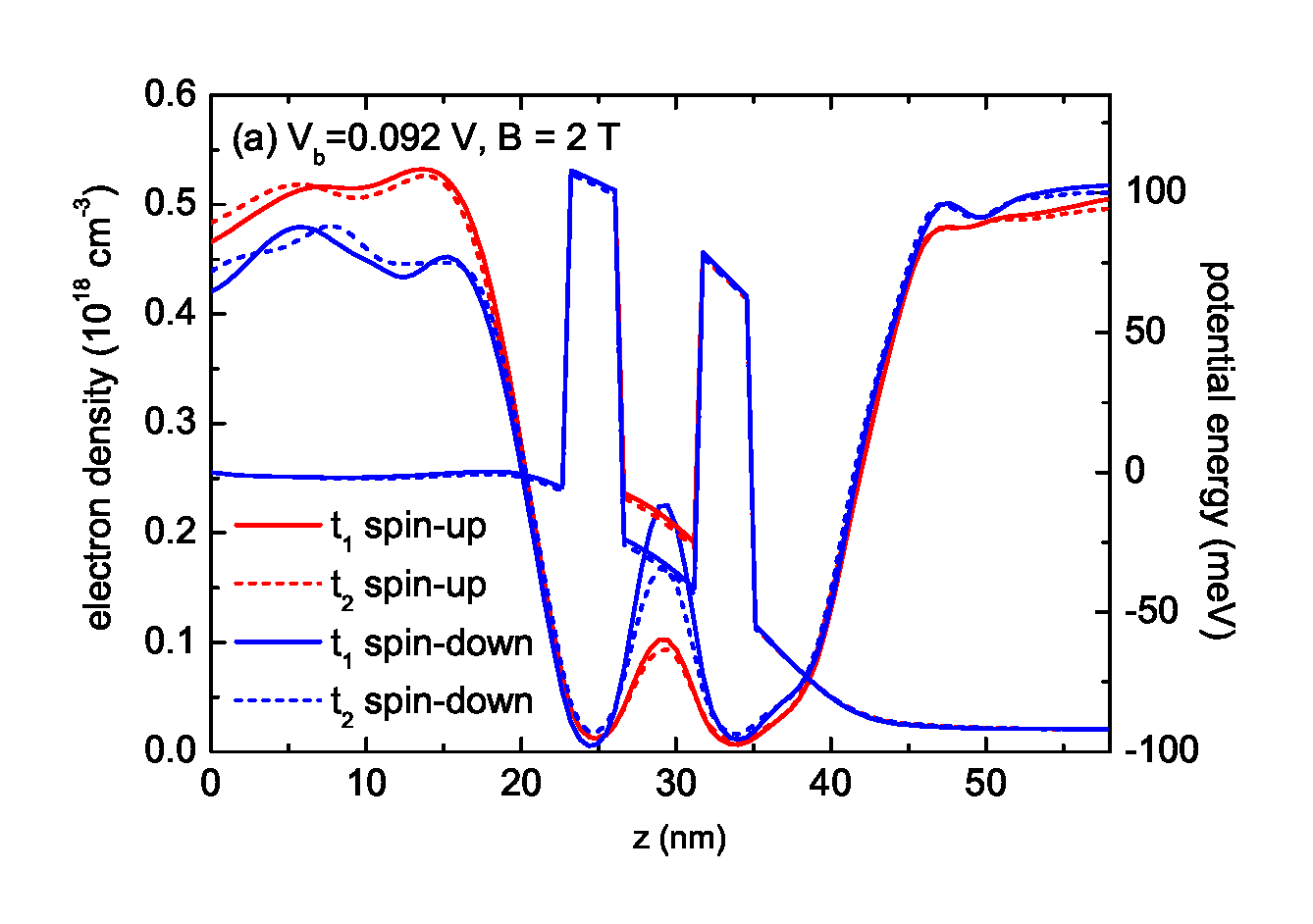}
\includegraphics[width=0.8\columnwidth]{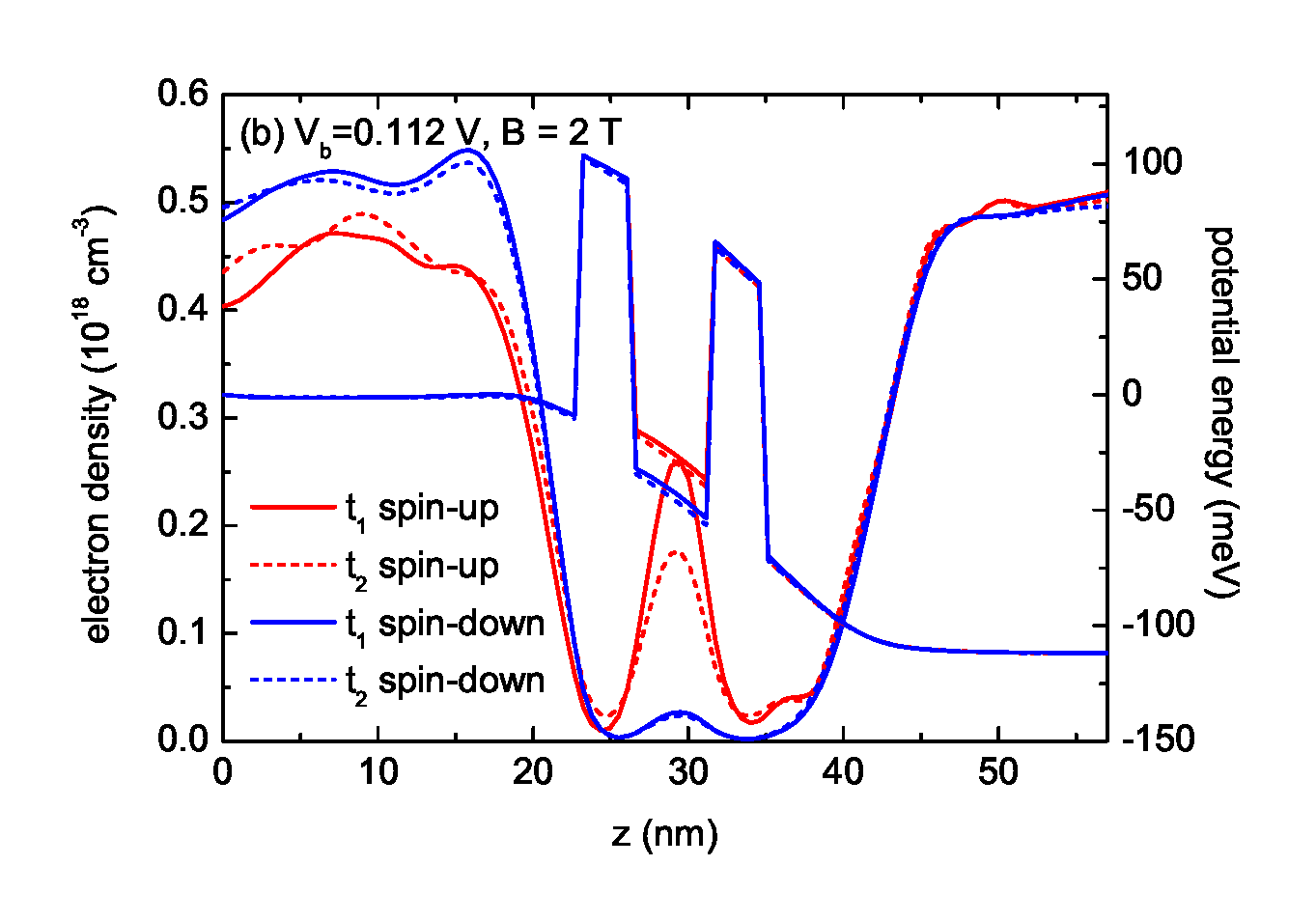}
\caption{\label{fig:7} (Color online) Density of the spin-up (red) and spin-down (blue) electrons
and the corresponding profiles of the self-consistent potential energy
as a function of coordinate $z$ at time instants $t_1$ and $t_2$
marked on Fig.~\ref{fig:6} for bias (a) $V_b=0.092$~V and (b) $V_b=0.112$~V.}
\end{figure}

In order to get a physical insight into the mechanism of oscillations of the spin-polarized currents,
we have calculated the time- and space-dependence of the electron density in the different regions of the
nanostructure (Figs.~\ref{fig:6} and \ref{fig:7}).
Figure~\ref{fig:6} shows the time dependence of spin-up and spin-down electron densities in the emitter region
and in the main quantum well (QW) calculated for the bias corresponding to the NDR regimes for spin-down
and spin-up current components.
We see that both the spin-up and spin-down electron densities oscillate with
the constant frequencies that are equal to the frequencies of the oscillations of the
corresponding spin-polarized current.
Figure~\ref{fig:6}(a) shows that in the NDR regime for the spin-down current component,
the amplitude of the spin-down electron density oscillations in the QW region is
much larger than the corresponding amplitude of the spin-up electron density oscillations.
At time instant $t_1$, the spin-down electron density  reaches the maximal value in the QW region
and the minimal value in the emitter region.
On the other hand, at time instant $t_2$, the spin-down electron density
is minimal in the QW and maximal in the emitter region [cf. Fig.~\ref{fig:6}(a)].
The similar behavior has been obtained for the spin-up electron density in the corresponding NDR regimes
for the spin-up current component [Fig.~\ref{fig:6}(b)].

Figure~\ref{fig:7} depicts the spatial distribution of the spin-up and spin-down electron densities
together with the corresponding self-consistent potential energy profiles determined at time instants $t_1$ and $t_2$
marked on Fig.~\ref{fig:6}.
Figure \ref{fig:7}(a) shows that the density of the spin-down electrons in the QW
is fairly large at time instant $t_1$ and much smaller at $t_2$.
These changes of the spin-down electron density in the QW
mean that the tunneling conditions at time instants $t_1$ and $t_2$ are different.
Therefore, the spin-down current component approaches (at $t_1$)  and partially leaves (at $t_2$)
the resonant tunneling conditions.
At these time instants, the density of the spin-up electrons in the QW is considerably smaller and
oscillates with smaller amplitude [Fig.~\ref{fig:7}(a)].
In this case, the low-amplitude oscillations of the spin-up electron density
are induced by the high-amplitude oscillations of the spin-down electron density
[cf. Fig.~(\ref{fig:6}(a)].
Similar results have been obtained for the spin-up electron density in the QW for the bias
corresponding to NDR regime of the spin-up current component [Fig.~\ref{fig:7}(b)].
At time instant $t_1$ the spin-up electron density in the QW is larger than at time instant
$t_2$. At these time instants, the density of the spin-down electrons in the QW is considerably
smaller and changes only slightly during the oscillations [Fig.~\ref{fig:7}(b)].
This indicates that the spin-down polarized current is out of the resonance, which explains
the small values of the spin-down current in the considered bias regime.

\section{Discussion}\label{sec:discusion}
The results of Sec.~\ref{sec:results} show that -- for the paramagnetic RTD in the external magnetic field --
we can expect two bias voltage regimes, in which the spin polarization of the current is not well defined but
oscillates with the THz frequency.
Figure \ref{fig:2} shows that the oscillations of the spin-polarized current are generated in
the two following bias ranges: the first corresponding to the NDR regime for the spin-down current component
and the second corresponding to the NDR regime for the spin-up current component.
The present results obtained for the paramagnetic RTD resemble those for the non-magnetic RTD,~\cite{Jensen1991}
according to which the current oscillations occur only in the NDR regime
of the current-voltage characteristics.
The previous studies~\cite{Orihashi2005,Kishimoto2007,Suzuki2009,Zhao2000,Zhao2001,Zhao2003}
of the current oscillations in the non-magnetic RTD allow us to
conclude that the occurrence of this phenomenon does not depend on the spins of
electrons flowing through the nanostructure.
In the next subsections, for simplicity and without loss of generality, first we
propose the model of the spin-independent current oscillations in the non-magnetic
RTD and next extend this model to the description of the spin-polarized
current oscillations in the paramagnetic RTD.

\subsection{Model of current oscillations}
In this subsection, we will analyze the current oscillations using the one-electron approximation.
In order to explain the origin of the oscillations, let us consider the physical processes occurring
in the NDR regime. In this regime, the system leaves the resonant tunneling conditions, which means that
the probability of the reflection of the electron from the emitter barrier becomes fairly large.
The interference between incident and reflected electron wave functions
together with the bias-induced lowering
of the electron potential energy leads to the formation of a shallow triangular quantum well in
the emitter region (cf. Fig.~\ref{fig:7}), in which the quasi-bound (resonance) state is formed.
Let us consider the coupling between the resonance state $\psi_1 \equiv \psi_E$ mainly localized in the
triangular quantum well in the emitter region and the resonance state $\psi_2 \equiv \psi_{QW}$ localized in the QW.
In the framework of the proposed model, we will calculate the probability of transition from state $\psi_1$
to state $\psi_2$ under influence of the weak perturbation corresponding to the change of the bias
and show that this probability oscillates in time.

The one-electron wave function $\Psi(z,t)$ evolves in time according to the time-dependent Schr\"odinger equation
\begin{equation}
\label{eq:nonSE}
i\hbar \frac{\partial}{\partial t} \Psi(z,t)=-\frac{\hbar ^2}{2m} \frac{\partial ^2}{\partial z^2} \Psi(z,t)+U[\Psi(z,t)]\Psi(z,t),
\end{equation}
where potential energy $U[\Psi(z,t)]$ is a functional of the wave function,
which results from the wave-function representation of the electron density
$n=n[\psi(z,t)]$ that in turn determines the potential energy
via Poisson equation (\ref{eq:PEq}).
Potential energy $U[\Psi(z,t)]$ can be expressed as the sum of three terms
\begin{equation}
\label{eq:nonU}
 U[\Psi(z,t)]=U^0(z)+\delta U_{V_b}(z)+ \delta U_H[\Psi(z,t)],
\end{equation}
where $U^0(z)$ is the potential energy of the electron in the steady state
for the bias $V_b=V^{res}_{b}$ that corresponds to the maximum of the resonant tunneling current.
The Hamiltonian with potential energy $U^0(z)$ describes the unperturbed system.
The next two terms in Eq.~(\ref{eq:nonU}) correspond to the perturbation, where $\delta U_{V_b}(z)$ is
the change of the electron potential energy caused by varying the bias from $V_b=V^{res}_{b}$
to $V_b=V_b^{res}+\delta V_b$ ($\delta V_b>0$) and
$\delta U_H[\Psi(z,t)]$ is the corresponding change of the Hartree potential energy.

The change of Hartree potential energy $\delta U_H[\Psi(z,t)]$ results from
changing the potential energy of the electron in the external electric field by
amount $\delta U_{V_b}(z)$ and occurs with some time delay
with respect to $\delta U_{V_b}(z)$.
Moreover, $\delta U_H[\Psi(z,t)]$ is much smaller than $\delta U_{V_b}(z)$, and
therefore, in the first approximation, it can be neglected
when considering the electron states at the bias $V_b=V_b^{res}+\delta V_b$.

In order to avoid the exponential divergence of the resonance wave function
in Eq.~(\ref{eq:nonSE}) all the calculations will be performed using the
complex-scaling theory of resonance states.~\cite{Moiseyev1998}
The complex-scaled Hamiltonian of the electron has the form
\begin{eqnarray}
 H(ze^{i\theta})&=&-\frac{e^{-2i\theta}\hbar ^2}{2m} \frac{\partial^2 }{\partial z^2}+U^0(ze^{i\theta})+\delta U_{V_b}(ze^{i\theta}) = \nonumber \\
&=&H_0(ze^{i\theta})+\delta U_{V_b}(ze^{i\theta})
\end{eqnarray}
where $\theta$ is the scaling parameter\cite{Moiseyev1998}
and $H_0(ze^{i\theta})$ is the Hamiltonian of the unperturbed system for the bias $V_b=V_b^{res}$.

In the NDR regime, i.e., for $V_b=V_b^{res}+\delta V_b$,
the resonance states $\psi^{\theta}_1(z)$ and $\psi^{\theta}_2(z)$
are almost degenerate, i.e., $E_1=E_2+\delta E$. Therefore, the one-electron wave function can be expressed as a linear
combination of the wave function $\psi^{\theta}_1(z)$
being localized in the emitter region and $\psi^{\theta}_{2}(z)$  localized in the QW
\begin{equation}
 \Psi^{\theta}(z,t)=a_1(t) \psi^{\theta}_{1}(z)+ a_2(t) \psi^{\theta}_{2}(z).
\label{eq:PSI}
\end{equation}
Wave functions $\psi^{\theta}_{n}(z)$ $(n=1,2)$
are calculated using the time-independent Schr{\"o}dinger equation for the unperturbed Hamiltonian
\begin{equation}
  H_0(ze^{i\theta})\psi^{\theta}_{n}(z)=E_n\psi^{\theta}_{n}(z),
\end{equation}
where the energy of the resonance state $E_n=\epsilon _n-i \Gamma _n /2$ consists of the real part $\epsilon _n$
that determines the position of the resonance on the energy scale and the imaginary part $\Gamma _n$ that defines the width of the resonance
and determines its lifetime.

After inserting (\ref{eq:PSI}) into the time-dependent Schr\"odinger equation (\ref{eq:nonSE}) we obtain
\begin{equation}
\label{eq:UR}
 i \hbar
\left (
\begin{array}{c}
 \frac{da_1(t)}{dt} \\
 \frac{da_2(t)}{dt}
\end{array}
\right ) =
\left (
\begin{array}{cc}
E_1 + W_{11} & W_{12} \\
W_{21} & E_2+W_{22}
\end{array}
\right )
\left (
\begin{array}{c}
a_1(t) \\
a_2(t)
\end{array}
\right ),
\end{equation}
where
\begin{equation}
 W_{nn^{\prime}}= \langle \psi_n^{\theta} | \delta U_{V_b} | \psi_{n^{\prime}}^{\theta} \rangle \;.
\end{equation}
Finally, the wave function $\Psi^{\theta}(z,t)$ takes on the form
\begin{eqnarray}
 \Psi^{\theta}(z,t)&=&\left ( a_1^0 e^{-\frac{i}{\hbar}\xi _+ t }
+ a_2^0 e^ {-\frac{i}{\hbar}\xi _- t} \right) \psi^{\theta}_1(z) \nonumber \\
&+&\left ( a_1^0 \alpha^+ e^ {-\frac{i}{\hbar}\xi _+ t }
+ a_2^0 \alpha ^- e^{-\frac{i}{\hbar}\xi _- t } \right) \psi^{\theta}_2(z),
\end{eqnarray}
where $a^0_{1,2}$ are the time-independent coefficients,
\begin{equation}
\alpha^{\pm} = \frac{W_{21}}{\xi_{\pm}-W_{22}} \;,
\end{equation}
\begin{equation}
 \xi_{\pm}  = \frac{1}{2} \left ( E_1+E_2+W_{11}+W_{22} \mp \sqrt{\Delta} \right ),  \\
\end{equation}
and
\begin{equation}
\Delta = (E_1+W_{11}-E_2-W_{22})^2+4|W_{12}|^2 \;.
\end{equation}
If we take the resonance state $\psi^{\theta}_1(z)$  as the initial state
of the system, i.e., the state for $t=t_0$ and $V_b=V_b^{res}$,
the probability of transition from state $\psi^{\theta}_1(z)$
localized in the emitter quantum well to state $\psi^{\theta}_{2}(z)$
localized in the QW is expressed by the formula
\begin{equation}
P_{1\rightarrow 2}(t)=\frac{4|W_{12}|^2}{\Delta}\sin^2 \omega t \;,
\label{eq:RE:1}
\end{equation}
where
\begin{equation}
\omega = \frac{\sqrt{\Delta}}{2 \hbar} \;.
\end{equation}
The probability of the electron transition from the QW to emitter is given by
\begin{equation}
P_{2\rightarrow1}(t) = 1-P_{1\rightarrow2}(t) \;.
\label{eq:RE:2}
\end{equation}
Based on Eqs.~(\ref{eq:RE:1}) and (\ref{eq:RE:2}) we state that the coupling
between the resonance states localized in the emitter and in the QW
(measured by $W_{12}$)
causes that the probabilities of the electron transition between the both
resonant states oscillate with frequency $\omega$.
As a consequence of this coupling, both the electron density and the current density oscillate
in certain well-defined bias intervals in the NDR regime.
Eqs.~(\ref{eq:RE:1}) and (\ref{eq:RE:2}) show that the densities
of electrons localized in the emitter and in the QW regions oscillate in anti-phase, which
supports the results presented in Fig.~\ref{fig:6}.

\subsection{Spin current oscillations in the paramagnetic RTD}
Based on the results of Sec. III and Subsection IV.~A
we argue that the oscillations of the spin-polarized currents in the NDR regimes
result from the coupling between the spin-dependent quasi-bound state localized in the
emitter region with the corresponding spin-dependent quasi-bound state localized in the main QW.

\begin{figure}[htbp]
\includegraphics[width=0.8\columnwidth]{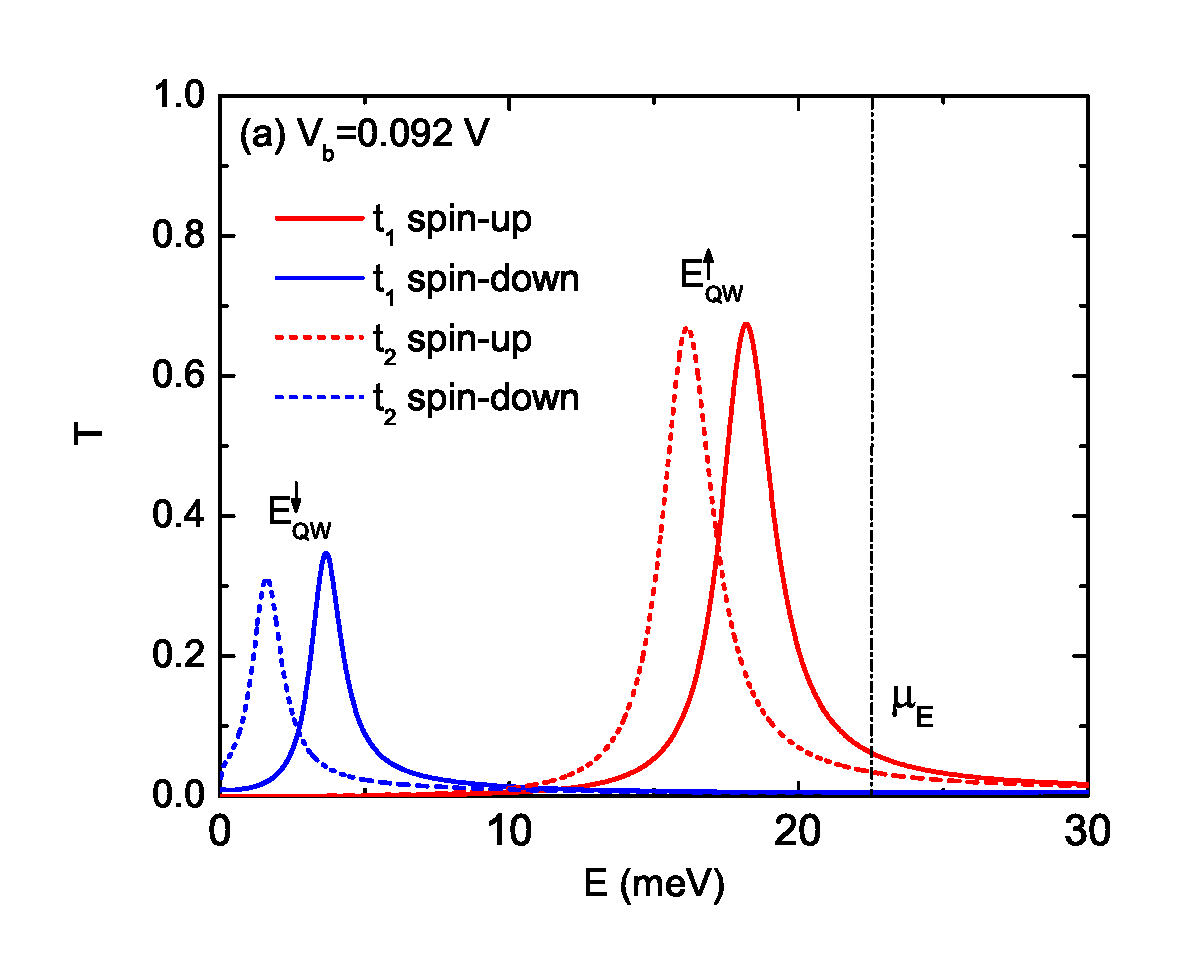}
\includegraphics[width=0.8\columnwidth]{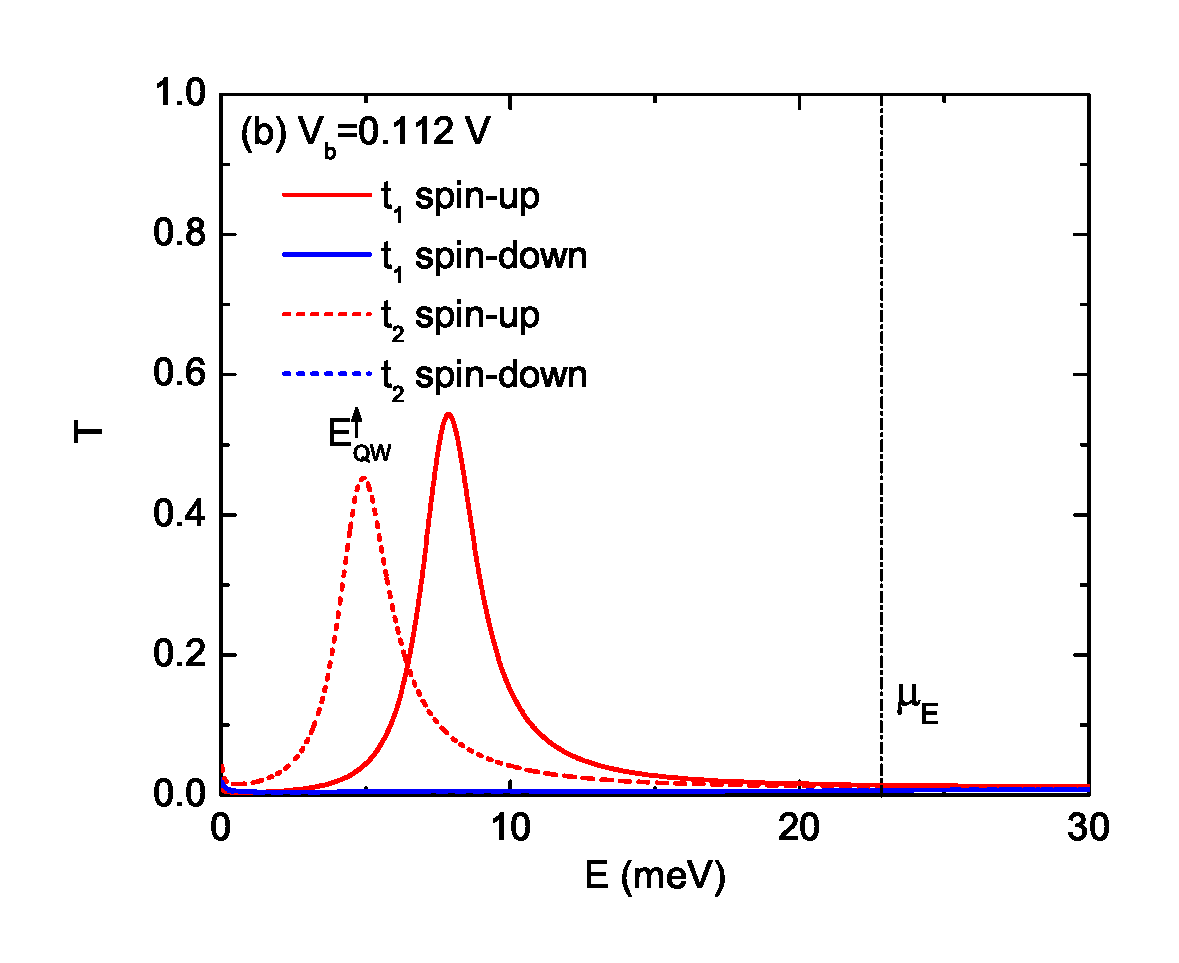}
\caption{\label{fig:8} (Color online) Transmission coefficient $T$ as a function of incident electron energy $E$
at time instants $t_1$ and $t_2$ marked on Fig.~\ref{fig:6} for the bias corresponding to the NDR regime for (a) spin-down
and (b) spin-up electrons.  The position of the peak determines the energy  of the
spin-up ($E_{QW}^{\uparrow}$, red line) and spin-down ($E_{QW}^{\downarrow}$, blue line) resonance state localized in the main QW.
$\mu_E$ is the electrochemical potential of the emitter.}
\end{figure}

Now we will analyze the formation of the oscillations of the spin-down polarized current
in the NRD regime of the current-voltage characteristics for this current component [cf. Fig.~\ref{fig:2}(a)].
In this regime, the resonant tunneling
conditions for the spin-down electrons are no longer satisfied and the reflection probability
of these electrons from the emitter barrier becomes fairly large.
Due to the interference and the bias-induced lowering of the potential energy,
the spin-down quasi-bound state of the electron is formed in the triangular quantum well
in the emitter region close to the left barrier [cf. Fig.~\ref{fig:7}(a)].
We have estimated the energy of this resonance state to be $E^{\downarrow}_E \simeq -0.1$~meV.
This energy differs by few meV from energy $E^{\downarrow}_{QW}$ of the resonance state
localized in the QW.  Therefore, these states can be treated as almost degenerate and
we can apply the results of Subsection IV.~A.
The coupling between these two quasi-bound states
leads to the oscillatory transitions of the spin-down electrons between the emitter and QW regions
[cf. Fig.~\ref{fig:6}(a)] described by the Eqs.~(\ref{eq:RE:1}) and (\ref{eq:RE:2}).
The oscillatory changes of the spin-down electron density in the QW
causes that the bottom of the QW oscillates in time, which gives rise to the
oscillatory changes of the tunneling conditions.
Figure~\ref{fig:8}(a) shows the transmission coefficient $T$ as a function of incident electron energy $E$ at time instants $t_1$ and $t_2$
marked on Fig.~\ref{fig:6}(a).
The positions of the peaks of $T(E)$ correspond to the energies of the
spin-up and spin-down resonance states localized in the main QW.
Figure~\ref{fig:8}(a) shows that at time instant $t_1$ the resonance tunneling condition is satisfied for the spin-down electrons,
which leads to the accumulation of these electrons in the QW [Fig.~\ref{fig:7}(a)].
After $\sim$1 ps  the system gets out of the resonance
and the density of the spin-down electrons in the QW decreases reaching the minimal value at
time instant $t_2$ [cf. Fig.~\ref{fig:7}(b)].
As a consequence, the spin-down polarized current oscillates [cf. Fig.~(\ref{fig:6}(a)].
For this bias voltage, the oscillations of the spin-up electron density are induced by the electron-electron interactions
with the oscillating spin-down electron density.

The same effect, i.e., the coupling between the two spin-up resonance states, one localized in the emitter quantum well
and the second localized in the QW, is responsible for the oscillations of the spin-up polarized current
that occur in the NRD regime for this current component [cf. Fig.~\ref{fig:2}(a)].
This coupling leads to the oscillatory exchange of the spin-up electrons between the emitter and the QW,
which changes the tunneling conditions in an oscillatory manner.

\subsection{Effect of exchange interaction}

The results presented in Sec. \ref{sec:results} have been
obtained without the exchange interaction [cf. Eq.~(\ref{Us})].
We have also studied the effect of exchange interaction
on the oscillations of spin polarized currents.
For this purpose we have estimated the exchange energy in
the QW using the formula\cite{exchange}
$E^{ex}_{\sigma} = - (3n_{\sigma}/\pi)^{1/3}/\varepsilon$,
which leads to $E^{ex}_{\sigma} \simeq -6$ meV
for the time instant, for which charge density $n_{\sigma}$
reaches the maximal value during the oscillation cycle.
Inserting this formula to Eq.~(\ref{Us}) we have calculated
the spin-dependent current-voltage characteristics and obtained
very similar results to those presented in Fig. \ref{fig:2}(a).
In particular, we have found two bias regimes, in which THz current oscillations
appear.  The estimated frequency of these oscillations is smaller
than that calculated without the exchange interaction, but still
remains in the THz range.  We conclude that the exchange interaction
can lead to a slight modification of the current-voltage characteristics,
but does not affect the predicted occurrence of the high-frequency spin current oscillations.

\section{Conclusions and Summary}\label{sec:concl}

In the present paper, we have shown that --
in the paramagnetic RTD under constant bias and external magnetic field-- the stable oscillations of the spin-polarized current
can occur in the NRD regimes for the spin-up and spin-down current components.
The analysis performed in Sec.~\ref{sec:discusion} reveals
that these oscillations result from the coupling between the spin-up (spin-down) resonance state localized in the emitter region
with the spin-up (spin-down) resonance state localized in the quantum well.
This coupling leads to the oscillatory changes of both the electron density and the current
in the nanostructure.

In the RTD under the fixed bias and external magnetic field, we deal with the following two types
of the spin-polarized currents: (i) stationary (constant in time) current with the well-defined
spin polarization, (ii) oscillating current with the oscillating spin polarization.
We have demonstrated that at a certain bias in the NDR regime, at which the spin polarization
of the net current reaches the maximum, the spin current polarization oscillates with the THz frequency.

Until now the THz oscillations of the current
have been observed only in non-magnetic resonant
tunneling diodes\cite{Orihashi2005,Kishimoto2007,Suzuki2009}
but they should be also occur in
paramagnetic resonant tunneling diodes.
We hope that the present results will stimulate the
experimental search for this  phenomena in
the magnetic resonant tunneling structures.
We would like to mention that the oscillations of the spin-polarized currents are fully manifested
only if the scattering on phonons and impurities is negligibly small, i.e., this is the low-temperature effect.
The scattering  processes will damp the oscillations and finally lead to the steady current.

The oscillations of the spin-polarized currents, predicted in the present paper,
can have important implications for spintronic devices based on the paramagnetic RTD.
The separation of both the spin components of the current is a basis for a fabrication of an effective spin filter (selector),
in which the spin polarization of the current is controlled by the bias voltage.
However, if the paramagnetic RTD is designed to operate as a spin filter, we have to avoid the bias regimes,
in which the current oscillations appear, in order to obtain the well-defined spin polarization
of the current.  On the other hand, the current oscillations obtained under the constant bias can be applied to
generate the spin-polarized current oscillating with the THz frequency.

In summary, we have demonstrated that the spin-polarized current flowing through the paramagnetic RTD
under the constant bias fixed in the NDR regime and in the constant external magnetic field can exhibit the THz oscillations.
The oscillatory behavior of the spin-polarized current leads to important consequences for the operation of spintronic devices.
On the one hand, this phenomenon is disadvantageous for the operation of spin filters as a parasitic effect
that spoils the operation of spin filters.  On the other hand,  it can be used to design the generator
of the spin-polarized currents oscillating with the THz frequency.

\begin{acknowledgments}
This paper has been supported by the Foundation for Polish Science MPD
Programme co-financed by the EU European Regional Development Fund
and the Polish Ministry of Science and Higher Education and its grants for
Scientific Research.
\end{acknowledgments}



%

\end{document}